\documentclass[preprintnumbers,twocolumn,amsmath,amssymb,floatfix,10pt,prd,
superscriptaddress,nofootinbib]{revtex4}
\usepackage{latexsym}
\usepackage{epsfig}
\usepackage{amssymb}

\begin{document}

\title{\large Gravitational lensing of wormholes in the galactic
   halo region}

\author{Peter K.F. Kuhfittig}
\email{kuhfitti@msoe.edu} \affiliation{Department of Mathematics,
Milwaukee School of Engineering, Milwaukee, Wisconsin 53202-3109,
USA}


\begin{abstract}
A recent study by Rahaman et al. has shown that the galactic halo
possesses the necessary properties for supporting traversable
wormholes, based on two observational results, the
Navarro-Frenk-White density profile and the observed flat
rotation curves of galaxies.  Using a method for calculating
the deflection angle pioneered by V. Bozza, it is shown that
the deflection angle diverges at the throat of the wormhole.
The resulting photon sphere has a radius of about 0.40 ly.
Given the dark-matter background, detection may be possible
from past data using ordinary light.\\
\phantom{a}


\end{abstract}

\maketitle

\section{Introduction}

It has been suggested quite recently that the galactic halo
possesses some of the characteristics needed to support
traversable wormholes \cite{RKRI}.  For such wormholes the
detection by means of gravitational lensing becomes a
distinct possibility, especially when viewed from a
strong-field perspective.

\section{Background}
It has been known for a long time that the rotation curves
of neutral hydrogen clouds in the outer regions of the
galactic halo cannot be explained in terms of ordinary
luminous matter.  The usual explanation is that galaxies
and even clusters of galaxies are pervaded by dark matter,
which does not emit electromagnetic waves nor interact
with normal matter.  To pursue the study of wormholes in
this context, we will rely on the Navarro-Frenk-White
density profile \cite{NFW96}:
\begin{equation}\label{E:rho}
\rho(r)=\frac{\rho_s}{\frac{r}{r_s}\left( 1+\frac{r}{r_s}
\right)^2},
\end{equation}
where $r_s$ is the characteristic scale radius and $\rho_s$
the corresponding density, to be described below.
Eq. (\ref{E:rho}) resulted from $N$-body simulations in the
search for the structure of dark halos in the standard CDM
cosmology.

One of the most important tools for the possible detection
of exotic objects such as wormholes is gravitational
lensing.  Since the exotic matter inside a wormhole
antigravitates, Cramer et al. \cite{Cramer95} and Safanova
\cite{STR02} examined the lensing effects of negative masses
on light rays from point sources.  Torres et al. \cite{TRA98}
suggested that wormholes can be probed using light curves of
gamma-ray bursts.

A method for calculating the gravitational microlensing
effect of the Ellis wormhole is derived in Ref. \cite{tA10}
and continued in Ref. \cite{TKAA10}.

More recently, gravitational lensing has been studied from
a strong-field perspective, often using an analytical
method pioneered by Bozza \cite{vB02} to calculate the
deflection angle.  This method was used in Refs.
\cite{TL05, TL12} to study two special models by Lemos
et al. \cite{jL03}.  Gravitational lensing by a $C$-field
wormhole is discussed in Ref. \cite{RKC07}.

\section{Wormhole structure}

As noted in the Introduction, the existence of dark matter
can be deduced from the observed flat rotation curves of
neutral hydrogen clouds in the outer regions of the halo.
The neutral hydrogen clouds must therefore be treated as
test particles moving in circular orbits.  Accordingly, the
spacetime in the galactic halo is characterized by the line
element \cite{RKRI}
\begin{equation}\label{E:line1}
ds^2=-B_0r^ldt^2+e^{2g(r)}dr^2
+r^2(d\theta^2+\sin^2\theta\,d\phi^2),
\end{equation}
using units in which $c=G=1.$  Here $l = 2(v^\phi)^2$,
where $v^{\phi}$ is the rotational velocity and $B_0$ is
an integration constant. (According to Ref. \cite{kN09},
$l\approx 0.000001$.)  In a wormhole setting a more
convenient form is \cite {MT88}
\begin{equation}\label{E:line2}
ds^2=-e^{2\Phi (r)}dt^2+\frac{dr^2}
{1-\frac{b(r)}{r}}
+r^2(d\theta^2+\sin^2\theta\,d\phi^2),
\end{equation}
where $e^{2\Phi (r)}=B_0r^{l}$.  So by letting
$B_0=1/b_0^{l}$, the spacetime metric becomes
\begin{equation}\label{E:line3}
ds^2=-\left(\frac{r}{b_0}\right)^ldt^2+\frac{dr^2}
{1-\frac{b(r)}{r}}
+r^2(d\theta^2+\sin^2\theta\,d\phi^2).
\end{equation}
From the Einstein field equation
\begin{equation}\label{E:Einstein}
   \frac{b'(r)}{r^2}=8\pi\rho(r),
\end{equation}
it is readily shown that
\begin{equation}\label{E:shape}
b(r)= 8 \pi \rho_s r_s^3 \left[\ln \left( 1+\frac{r}{r_s} \right)
+ \frac{1}{ 1+\frac{r}{r_s} }+C \right],
\end{equation}
where $C$ is an integration constant.  We now recall that
if Eq. (\ref{E:line2}) represents a wormhole, then\\
(1) The \emph{redshift function}, $\Phi (r)$, must remain finite
to prevent an event horizon. According to Eq. (\ref{E:line3}),
this condition is met.  It also follows that the wormhole
spacetime is not asymptotically flat.\\
(2) The \emph{shape function}, $b(r)$, must obey the following
conditions at the throat $r = r_{th}$ :  $b(r_{th}) = r_{th}$
and $b^\prime(r_{th}) < 1$, the so-called flare-out condition.
Moreover, Eq. (\ref{E:Einstein}) implies that $b'(r)$ must be
positive.\\
(3) For $b=b(r)$ we also have $b(r)<r$ for $r >r_{th}$.\\
Finally, the expressions for the radial pressure, $p_r$, and
the lateral pressure, $p_t$, are given in Ref. \cite{RKRI}.
It is also shown that $\rho+p_r<0$ near the throat.

Before analyzing the shape function, we return to Ref.
\cite{NFW96} to note some of the proposed forms of $\rho(r)$:
\begin{equation*}
  \rho(r)\propto \frac{1}{r\left(1+\frac{r}{r_s}\right)^3}
  \quad \text{and} \quad
   \rho(r)\propto \frac{1}{r\left(1+\frac{r}{r_s}\right)^2}.
\end{equation*}
These forms suggest a more general starting point:
\begin{equation}\label{E:start}
  \rho(r)= \frac{K}{r\left(1+\frac{r}{r_s}\right)^n},
  \quad n>1,
\end{equation}
for some $K>0$.  This form yields
\begin{equation}
  b(r)=\frac{-8\pi Kr_s^2}{(n-1)(n-2)}\frac{(n-1)
  \frac{r}{r_s}+1}{\left(1+\frac{r}{r_s}\right)^{n-1}}+C
\end{equation}
and
\begin{equation}
  b'(r)=\frac{8\pi Kr_s}{2-n}\frac{1-\frac{(n-1)r/r_s+1}
  {1+r/r_s}}
  {\left(1+\frac{r}{r_s}\right)^{n-1}},\quad n\neq 2.
\end{equation}
We now see that $b'(r)$ reduces to
\begin{equation}\label{E:der}
    b'(r)=8\pi Kr_s\frac{\frac{r}{r_s}}
    {\left(1+\frac{r}{r_s}\right)^n},
\end{equation}
which is valid for $n=2$, corresponding to
Eq. (\ref{E:rho}).
We also have $b'(r)<1$ for all $r$, provided that
$K<8\pi r_s$.  This condition is easily met
for all the cases discussed in TABLE 1 of Ref.
\cite{NFW96}: $r_s$ is defined in terms of the ``virial"
radius, which ranges from 177 kpc for a dwarf galaxy
to 3740 kpc for a rich galaxy cluster.  [See Ref.
\cite{NFW96} for details.]  That $b'(r)<1$ near the
throat can also be seen from Fig. 1.

Returning now to the shape function, Eq. (\ref{E:shape}),
one way to determine the radius of the throat is to
define the function $B(r)=b(r)-r$ and locate the root
$r=r_{th}$ (if it exists) of $B(r)=b(r)-r=0$. In other
words, we are treating $B(r)$ as if it were a function
in rectangular coordinates.  This approach is possible
because of the spherical symmetry: we can move radially
outward in any direction, thereby forming the $r$-axis.
These facts will be exploited further in Sec.
\ref{S:location}.  Next, we observe that since $B'(r)=
b'(r)-1$, the condition $b'(r)<1$ near the throat and
$b'(r)>0$ imply that
\begin{equation}\label{E:Bprime}
    -1<B'(r)<0.
\end{equation}
So $b(r)-r$ is strictly decreasing.  If $b(r)-r$
does not intersect the $r$-axis, it must have a
horizontal asymptote, so that
$\text{lim}_{r\rightarrow \infty}(b'(r)-1)=0$.  But
this is impossible since
$\text{lim}_{r\rightarrow\infty}b'(r)=0$ by Eq.
(\ref{E:der}).  So for some
$r=r_{th}$, $b(r_{th})=r_{th}$, which is the
throat of the wormhole.  We will see in the next
section that this conclusion can be reached more
easily from the graph of $b=b(r)$, thereby
yielding an alternative to the method in
Ref. \cite{RKRI}.

\section{Further discussion of $\rho(r)$}
\label{S:rho}

Since the discussion of our wormhole structure
depends on Eq. (\ref{E:rho}), we need to take a
closer look at the parameters used, as well as the
coordinate system.  In particular, a more complete
description of the density is
\begin{equation}\label{E:rho2}
   \rho(r)=\frac{\delta_c\rho_{crit}}{\frac{r}{r_s}
   \left(1+\frac{r}{r_s}\right)^2}.
\end{equation}
Here $\rho_{crit}$ is the critical density that is
given by $\rho_{crit}=3H^2/8\pi G$, where $H$ is the
current value of the Hubble constant.  The other
parameter is
\begin{equation*}
  \delta_c=\frac{200}{3}\frac{c^3}{\text{ln}\,(1+c)
  -\frac{c}{1+c}}
\end{equation*}
with the incorporated concentration parameter
\[
      c=\frac{r_{200}}{r_s},
\]
which is defined in terms of two additional
parameters, $r_{200}$, the ``virial" radius and
$r_s$, a characteristic scale radius.  [See Ref.
\cite{NFW96} for details.]  As already noted,
TABLE 1 in Ref. \cite{NFW96} lists the values
of these parameters for different systems
ranging from dwarf galaxies to rich galactic
clusters.  Being primarily interested in our
own galaxy, we will use the values in Line 5
of TABLE 1, i.e.,
$r_s/r_{200} =0.060$, $r_{200}=348\,\text{kpc}$,
resulting in $r_s=20.88$ kpc.

For the time being we need to assume that the
center of our wormhole is located at the origin
$O$.  This is described in Ref. \cite{NFW96} as
the center of the halo, which, in turn, is the
center of mass of certain ``clumps" \cite{NFW96}.
Other possible locations of these wormholes are
discussed in Sec. \ref{S:location}.

Returning once again to the shape function, Eq.
(\ref{E:shape}), observe that, qualitatively,
$b(r)$ has the form
\begin{equation}\label{E:form1}
   b(r)=A\left[\text{ln}\,\left(1+\frac{r}{B}\right)
   +\frac{1}{1+\frac{r}{B}}+C\right].
\end{equation}
Fig. 1 shows that a throat of radius $r=r_{th}$ will
\begin{figure}[htbp]
        \includegraphics[scale=.6]{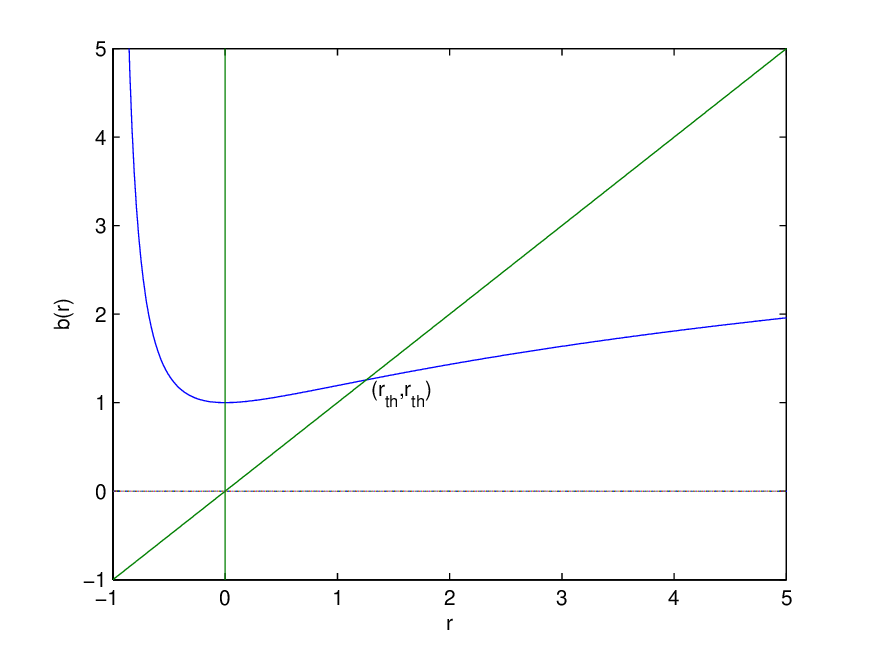}
        \caption{Qualitative features of the shape
        function showing the existence and position
        of the throat.}
     \label{fig:shape2}
\end{figure}
always exist.  However, based on our model, there
is no way to determine its size from purely
geometric considerations.  In other words,
additional information is needed and this comes
from the gravitational lensing discussed in the
next section.

\section{Gravitational lensing}

As noted in the Introduction, Bozza \cite{vB02}
provided an analytic method for calculating the
deflection angle for any spherically symmetric
spacetime in the strong field limit.  To apply
these methods to wormholes, we will follow the
procedure in Refs. \cite{TL05, TL12}.

As in most discussions on gravitational lensing,
the line element is taken to be
\begin{equation}
ds^2=-A(x)dt^2+B(x)dx^2
+C(x)(d\theta^2+\sin^2\theta\,d\phi^2),
\end{equation}
where $x$ is the radial distance defined in terms
of the Schwarzschild radius $x=r/2M$.  Then
\begin{equation}\label{E:closest}
   x_0=\frac{r_0}{2M}
\end{equation}
denotes the closest approach of the light ray.

Using the lens equation in Ref. \cite{VE00}, it is
shown in Refs. \cite{TL05, TL12} that the
deflection angle $\alpha(x_0)$ consists of the
sum of two terms:
\begin{equation}\label{E:angle1}
  \alpha(x_0)=\alpha_e+I(x_0).
\end{equation}
Here
\begin{equation}\label{E:angle2}
   \alpha_e=-2\, \text{ln}\left(\frac{2a}{3}-1\right)
   -0.8056
\end{equation}
is due to the external Schwarzschild metric outside
the wormhole's mouth $r=a$; $I(x_0)$ is the
contribution from the internal metric, given by
\begin{equation}\label{E:deflection1}
    I(x_0)=2\int^{\infty}_{x_0}\frac{\sqrt{B(x)}\,dx}
    {\sqrt{C(x)}\sqrt{\frac{C(x)A(x_0)}{C(x_0)A(x)}-1}}.
\end{equation}

From our line element (\ref{E:line3}) and the shape
function (\ref{E:shape}), we obtain
\begin{equation}\label{E:deflection2}
   I(x_0)=\\ \int^a_{x_0}Q(x)\,dx,
\end{equation}
where
\begin{multline}
   Q(x)=\\ \frac{2}{\sqrt{x^2\left\{1-
   \frac{k}{2M}\frac{1}{x}\left[\text{ln}\,
   \left(1+\frac{x}{x_s}\right)
   +\frac{1}{1+\frac{x}{x_s}}+C\right]\right\}}
   \sqrt{\frac{x^{2-l}}{x_0^{2-l}}-1}}.
\end{multline}
Here $k=8\pi\rho_sr_s^3$ from Eq. (\ref{E:shape}).
Observe that $\rho_sr_s^2$ is dimensionless, so
that $\rho_sr_s^2=\rho_sx_s^2$.  It follows that
$k$ has units of length; hence, in Schwarzschild
units,
\begin{equation}\label{E:k}
  \frac{k}{2M}=8 \pi \rho_s x_s^2
  \frac{r_s}{2M}=8 \pi \rho_s x_s^3.
\end{equation}
  To see where this integral diverges, we make the
change of variable $y=x/x_0$:
\begin{equation}\label{E:deflection3}
    I(x_0)=\int_1^{a/x_0}R(y)\,dy,
\end{equation}
where
\begin{multline}
   R(y)=\\ \frac{2}{\sqrt{(y^{4-l}-y^2)
    \left\{1-\frac{k}{2Mx_0}\frac{1}{y}\left[\text{ln}
    \left(1+\frac{yx_0}{x_s}\right)+
    \frac{1}{1+\frac{yx_0}{x_s}}+C\right]\right\}}}.
\end{multline}
The radicand $F(y)$ in the denominator can be expanded
in a Taylor series around $y=1$.  Letting
\begin{equation}\label{E:form2}
    g(y)=1-\frac{k/2M}{x_0}\frac{1}{y}\left[\text{ln}
    \left(1+\frac{yx_0}{x_s}\right)+
    \frac{1}{1+\frac{yx_0}{x_s}}+C\right],
\end{equation}
we obtain
\begin{multline}
   F(y)=(2-l)g(1)(y-1)\\+
   \left[\frac{1}{2}(5-l)(2-l)g(1)+(2-l)g'(1)\right]
   (y-1)^2\\+
    \text{higher powers}.
\end{multline}
If $g(1)\neq 0$, the integral converges due to the leading
term $(y-1)^{1/2}$ resulting from the integration.  If
$g(1)=0$, then the second term leads to $\text{ln}\,(y-1)$,
which causes the integral to diverge.  (As an illustration,
the integral
\begin{equation*}
   \int_1^2\frac{dx}{\sqrt{3(x-1)+2(x-1)^2+(x-1)^3+
   5(x-1)^4}}
\end{equation*}
converges, but the integral
\begin{equation*}
   \int_1^2\frac{dx}{\sqrt{2(x-1)^2+(x-1)^3+
   5(x-1)^4}}
\end{equation*}
does not.)

So we need to examine $g(y)$ more closely.
Observe that from Fig. 1, for any $y_m>0$, there
exists a constant $C_1$ such that
\begin{equation}\label{E:form3}
\text{ln}\left(1+\frac{y_mx_0}{x_s}\right)+
    \frac{1}{1+\frac{y_mx_0}{x_s}}+C_1=y_m.
\end{equation}
[The form of the function remains that of Eq.
(\ref{E:form1}).]
Hence from Eq. (\ref{E:form2}),
\[
    1-\frac{k/2M}{x_0}\frac{1}{y_m}(y_m)=0
\]
and $x_0=k/2M$.  So for some $C_2$, i.e., translating
in the vertical direction,
\begin{multline}\label{E:form4}
    g(1)=\\1-\frac{k/2M}{x_0\times 1}\left[\text{ln}
    \left(1+\frac{1\times x_0}{x_s}\right)+
    \frac{1}{1+\frac{1\times x_0}{x_s}}+C_2\right]
    =0,
\end{multline}
and again $x_0=k/2M$.  In this manner we have
obtained the value of the closest approach
in Schwarzschild units.  By reinterpreting
Eq. (\ref{E:form4}), i.e.,
\begin{multline}\label{E:form5}
    g(1)=\\1-\frac{1}{2Mx_0}\left\{k\left[\text{ln}
    \left(1+\frac{x_0}{x_s}\right)+
    \frac{1}{1+\frac{x_0}{x_s}}+C_2\right]\right\}=\\
    1-\frac{1}{r_0}\left\{k\left[\text{ln}
    \left(1+\frac{r_0}{r_s}\right)+
    \frac{1}{1+\frac{r_0}{r_s}}+C_2\right]\right\}=0,
\end{multline}
we see that
\[
    1-\frac{b(r_0)}{r_0}=0 \quad \text{and} \quad
    b(r_0)=r_0.
\]
So $r_0$, the closest approach, coincides with the
throat, i.e., $r_0=r_{th}$.

According to Refs. \cite{TL05, TL12}, this is also
the radius of the photon sphere.  It remains to
determine the value of $k$.  From Eq. (\ref{E:k}),
\[
  \frac{k}{2M}=8\pi\rho_sx_s^2\frac{r_s}{2M}=x_0
  =\frac{r_0}{2M},
\]
and we may revert to $k=8\pi\rho_sr_s^3=r_0=r_{th}$.  We
now find that
\[
    r_{th}=3.7466\times 10^{15}\, \text{m}\approx
    0.40\, \text{ly}.
\]

\section{The location of the wormhole}\label{S:location}

We saw earlier that the throat is the $r$-intercept of
$B(r)=b(r)-r$ since, due to the spherical symmetry,
$B(r)$ is the same function along any outward ray
emanating from the origin.  In similar manner, if a
shape function is viewed as a translation to or from
the origin along this ray, this would correspond to a
horizontal translation in Fig. 1.  Upon closer
examination, however, such a translation would affect
the radius of the throat.  But recalling that $b(r)$
has the form in Eq.(\ref{E:form1}), namely,
 \begin{equation*}
   b(r)=A\left[\text{ln}\,\left(1+\frac{r}{B}\right)
   +\frac{1}{1+\frac{r}{B}}+C\right],
\end{equation*}
a larger $B$ will cause the curve to ``flatten," as
shown in Fig. 2.  Since $r_s\approx 20.88$ kpc is
\begin{figure}[htbp]
        \includegraphics[scale=.6]{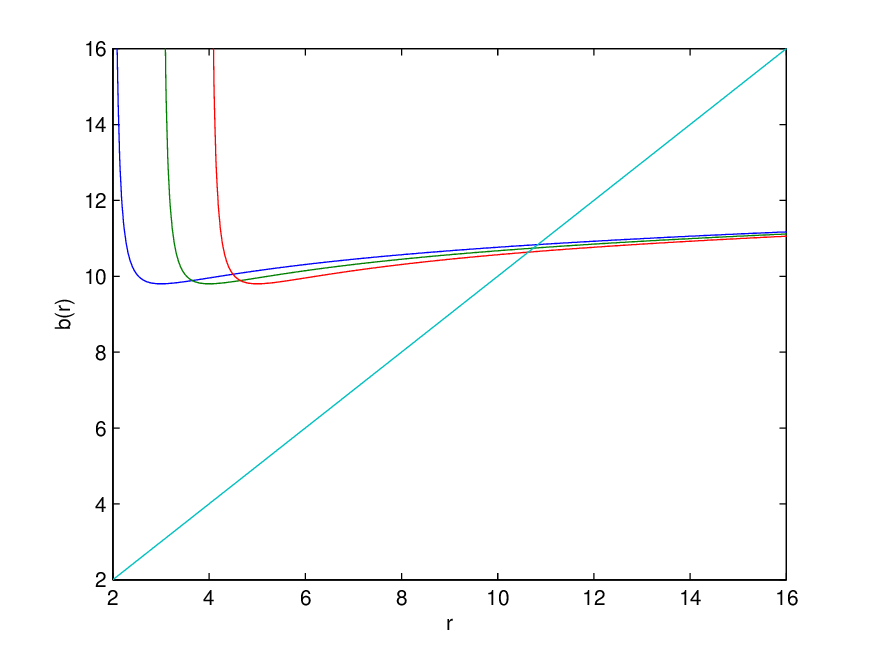}
        \caption{A horizontal translation of $b(r)$ has a
        negligible effect on the throat size.}
   \label{fig:shape2}
\end{figure}
indeed very large compared to $r$ not too far from
the origin $O$, the effect of the translation on
the throat size is small or even negligible.
Suppose that the center of the wormhole is at $O'$,
the new origin (Fig. 3).  Then the throat
radius relative to $O'$
\begin{figure}[htbp]
        \includegraphics[scale=.6]{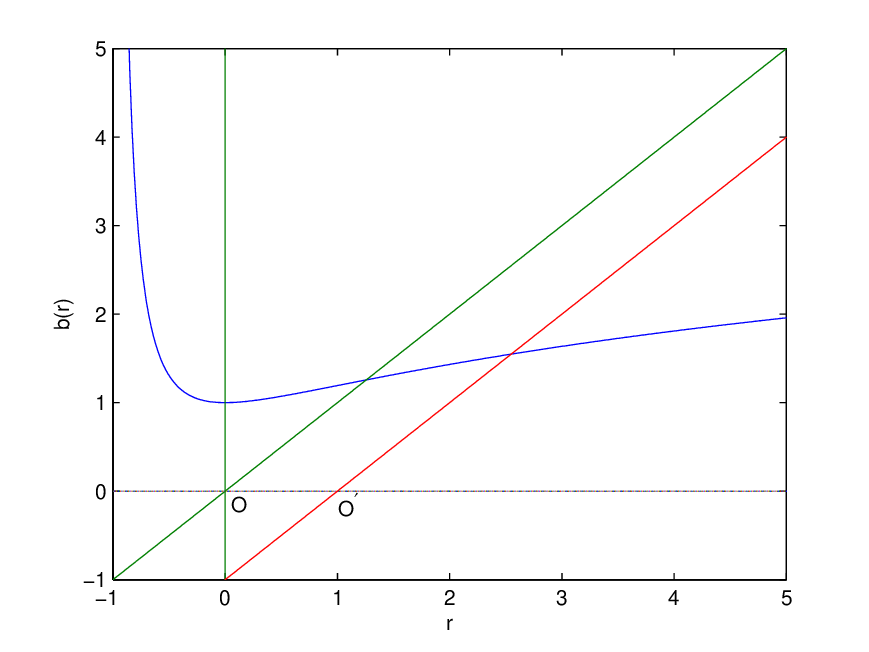}
        \caption{A wormhole centered at $O$ cannot
        be distinguished from a wormhole centered
        at $O'$.}
   \label{fig:shape2}
\end{figure}
is approximately the same as in Fig. 1.  So
mathematically speaking, $b(r)$ has the same
properties regardless of the location along the
ray.  In other words, mathematically, and hence
physically, we cannot distinguish between a
wormhole centered at $O$ from a wormhole
centered at $O'$.  In particular, the radius of the
throat is approximately 0.40 ly for all wormholes
in our model, provided, of course, that they exist.
Since the radius of the throat corresponds to
the radius of the photon sphere, such a sphere
would be detectable, thereby providing
observational evidence for the existence of
the wormhole.

To obtain further evidence, it would be highly
desirable to obtain at least a rough estimate
of the radius $r=a$ of the mouth of the wormhole.
Here we return to the origin $O$.  As noted earlier,
the wormhole spacetime is not asymptotically flat,
raising the question of a possible cutoff at $r=a$
and the junction to an external Schwarzschild
spacetime.  Here we are going to assume that even
though we are still in the halo region, the
exterior solution is close enough to Schwarzschild
to permit a rough estimate of $r=a$.  To that end
we recall that the mass of the wormhole is $M=
\frac{1}{2}b(a)$, where $a<r_s$.  Using the
redshift function in line element (\ref{E:line3}),
$a$ is determined from the condition $(a/b_0)^l
=1-2M/a$, i.e.,
\begin{equation}\label{E:junction}
   \left(\frac{a}{b_0}\right)^{l}=1-
   \frac{k\left[\text{ln}\left(1+\frac{a}{r_s}
   \right)+\frac{1}{1+\frac{a}{r_s}}+C\right]
   }{a},
\end{equation}
where $b_0$ is a constant of integration.

Before continuing, we need to estimate the value
of $C$, given the earlier parameters.  Since $k=
r_{th}$ and $b(r_{th})=r_{th}$, we have from the
shape function
\[
    b(r)=k\left[\text{ln}\left(1+\frac{r}{r_s}
    \right) +\frac{1}{1+\frac{r}{r_s}}+C\right]
\]
that
\[
    \text{ln}\left(1+\frac{r_{th}}{r_s}\right)
    +\frac{1}{1+\frac{r_{th}}{r_s}}+C=1,
\]
which leads to $C\approx -1.7\times 10^{-11}$.
It follows that $C$ can be safely neglected in
Eq. (\ref{E:junction}).

A return to the parameters discussed in
Sec. \ref{S:rho} now offers some posibilities:
suppose we let $b_0=r_{200}=348\,\text{kpc}$,
the virial radius, and then use
$r_s=20.88\,\text{kpc}=6.44\times
10^{20}\, \text{m}$, $l=0.000001$, and $k=
3.7466\times 10^{15}\,\text{m}$.  These
values suggest that we should expect
$r=a$ to be in excess of 1 $r_s$.  Evidence
for the existence of a mouth can therefore
be sought in this range.


It has been suggested that for wormholes in our
halo, the Large Magellanic Cloud can be used for
applying the gravitational lensing technique,
possibly even using past data.  For the location
of the mouth, we need to recall that according to
Refs. \cite{TL05, TL12}, photons with closest
approach greater than the wormhole's mouth have
a Schwarzschild lensing effect.  For both throat
and mouth, these effects can in principle be
detected.
\\
\\
\emph{Remark:}  While it is known that the NFW
model predicts velocities in the central parts
that are too low \cite{gG14}, these discrepancies
vanish in the more remote regions of the halo
\cite{cT06, aM12}.


\section{Summary}

Given the Navarro-Frenk-White density profile of
halos, it was shown in Ref. \cite{RKRI} (and
confirmed in this paper) that these halos posses
some of the characteristics that could give rise
to traversable wormholes. 
The shape function, obtained from this profile, meets
the flare-out condition at the throat.  Such a throat
always exists, but its location cannot be determined
from the geometry.  However, using a method for
calculating the deflection angle pioneered by Bozza
\cite{vB02}, it is shown that the deflection angle
diverges at the throat.  The resulting photon sphere
has a radius of about 0.40 ly regardless of the
location of the wormhole.  Detection may be possible
using past data.  Since the dark matter in the halo
region does not interact with light, a suitable
vehicle would be ordinary light from the Large
Magellanic Cloud.


\begin{thebibliography}{99}

\bibitem{RKRI}F. Rahaman, P.K.F. Kuhfittig, S. Ray, and
   N. Islam, Eur. Phys. J. C \textbf{74}, 2750 (2014).
\bibitem{NFW96}J.F. Navarro, C.S. Frenk, and S.D.M. White,
   Astroph. J. \textbf{462}, 563 (1996).
\bibitem{Cramer95}J.G. Cramer, R.L. Forward, M.S. Morris,
   M. Visser, G. Benford, and G.A. Landis, Phys. Rev. D
   \textbf{51}, 3117 (1995).
\bibitem{STR02}M. Safanova, D.F. Torres, and G.E. Romero,
   Phys. Rev. D \textbf{65}, 023001 (2002).
\bibitem{TRA98}D.F. Torres, G.E. Romero, and L.A.
   Anchordoqui, Phys. Rev. D \textbf{58}, 123001 (1998).
\bibitem{tA10}F. Abe, Astrophys. J. \textbf{725},
     787 (2010).
\bibitem{TKAA10}Y. Toki, T. Kitamura, H. Asada, and
   F. Abe, Astrophys. J. \textbf{740}, 121 (2010).
\bibitem{vB02}V. Bozza, Phys. Rev D \textbf{66}, 103002
   (2002).
\bibitem{TL05}J.M. Tejeiro and E.A. Larranaga, arXiv:
   gr-qc/0505054.
\bibitem{TL12}J.M. Tejeiro and E.A. Larranaga, Rom. J.
   Phys. \textbf{57}, 736 (2012).
\bibitem{jL03}J.P.S. Lemos, F.S.N. Lobo, and S. Quinet
   de Oliveira, Phys. Rev. D \textbf{68}, 064004 (2003).
\bibitem{RKC07}F. Rahaman, M. Kalam, and S. Chakraborty,
   Chin. J. Phys. \textbf{45}, 518 (2007).
\bibitem{kN09}K.K. Nandi, A.I. Filippov, F. Rahaman,
  S. Ray, A.A. Usmani, M. Kalam, and A. DeBenedictis,
  Mon. Not. Roy. Astron. Soc. \textbf{399}, 2079 (2009).
\bibitem{MT88} M.S. Morris and K.S. Thorne, Am. J. Phys.
   \textbf{56}, 395 (1988).
\bibitem{VE00}K.S. Virbhadra and G.F.R. Ellis, Phys.
    Rev D \textbf{62}, 084003 (2002).
\bibitem{gG14}G. Gentile, P. Salucci, U. Klein, D.
   Vergani, and P. Kalberla, Mon. Not. Roy. Astron.
   Soc. \textbf{351}, 903 (2014).
\bibitem{cT06}C. Tonini, A. Lapi, and P. Salucci,
   Astroph. J. \textbf{649}, 591 (2006).
\bibitem{aM12}A.V. Maccio et al., ApJ Lett.
   \textbf{744}, L9 (2012).

\end{thebibliography}
\end{document}